\documentclass[epjST]{svjour}
\usepackage{graphicx}
\usepackage{graphics}
\usepackage{amsmath}
\usepackage{amsfonts}
\usepackage{amssymb}
\begin{document}
\title{On the symplectic integration of the discrete nonlinear Schr\"{o}dinger equation with disorder}
\author{Enrico Gerlach \inst{1}\fnmsep\thanks{\email{enrico.gerlach@tu-dresden.de}} \and Jan Meichsner\inst{1} \and Charalampos Skokos\inst{2}}
%
\institute{Lohrmann Observatory,
  Technical University Dresden, D-01062, Dresden, Germany \and Department of
    Mathematics and Applied Mathematics, University of Cape Town,
    \mbox{Rondebosch}, 7701, South Africa}
\abstract{
We present several methods, which utilize symplectic integration techniques  based on two
and three part operator splitting, for numerically solving the equations
of motion of the disordered, discrete nonlinear Schr\"{o}dinger (DDNLS)
equation, and compare their efficiency. Our results suggest that the most
suitable methods for the very long time integration of this one-dimensional Hamiltonian
lattice model with many degrees of freedom (of the order of a few hundreds) are the ones
based on three part splits of the system's Hamiltonian. Two part split techniques
can be preferred for relatively small lattices having up to $N\approx\;$70
sites. An advantage of the latter methods is the better conservation of the
system's second integral, i.e.~the wave packet's norm.
} 
\maketitle
%

\section{Introduction}
\label{sec:intro}

Disordered systems are  models of usually many degrees of freedom trying to mimic heterogeneity in nature. Typically they are obtained by attributing to one of the system's parameters a different, random value for each degree of freedom. It is well-known that in linear disordered systems energy excitations remain localized. This phenomenon was first studied by Anderson in 1958 \cite{A58}  and for this reason is usually called `Anderson localization'. This behavior plays an important role in several physical processes, like for example the conductivity of materials, the dynamics of Bose-Einstein condensates etc.

In the last decade the effect of nonlinearity on disordered systems has
attracted extensive attention in theory and simulations
\cite{VKF09_B11_MAP11_MAPS11_LBF12_MP12_MP13_LIF14,KKFA08,PS08,FKS09,GMS09,SKKF09,SF10,LBKSF10,BLSKF11,SGF13,ABSD14}, as well as in experiments \cite{SBFS07_LAPSMCS08_REFFFZMMI08}.
A fundamental question in this context is what happens to energy localization in the presence of nonlinearities. Several studies of  the disordered variants of two fundamental one-dimensional Hamiltonian lattice models, namely the Klein-Gordon (KG) oscillator chain and the discrete nonlinear Schr\"{o}dinger equation (DDNLS), determined the statistical characteristics of energy spreading and showed that nonlinearity destroys localization \cite{FKS09,GMS09,SF10,LBKSF10,BLSKF11}. In these works the existence of different dynamical behaviors and spreading regimes was revealed, their particular dynamical characteristics were determined and their appearance was theoretically explained. The DDNLS model
was used for the theoretical treatment of wave packet spreading,
while numerics in both models were employed for verifying the obtained
theoretical results.

The numerical integration of the
KG model proved to be computationally easier than the DDNLS system, as the
KG Hamiltonian can be readily split into
two integrable parts (namely the kinetic and the
potential energy). This splitting  permits  the application of several commonly used symplectic integrators (SIs) for the integration of the KG model, like for
example the SABA$_2$ integrator with corrector \cite{LR01},
which proved to be a very efficient fourth order SI
for this system \cite{FKS09,SKKF09,SF10,LBKSF10,BLSKF11,SGF13}.
On the other hand, the numerical integration of the DDNLS
model is a more complicated task, which
requires the implementation of some specially designed techniques (see for example the
appendices of \cite{SKKF09,BLSKF11} for more details). Nevertheless, the required CPU times
did not allow the computation of the system's evolution up to the same final
times as for the KG model; typically for the same computational effort wave packets in the KG model
were propagated in time up to one or two orders of magnitude longer than in the DDNLS system.

Although, nowadays it is common knowledge that energy spreading in disordered lattices is a chaotic process, the characteristics of this chaotic behavior have not been studied in detail. The first attempt to systematically investigate chaos in one-dimensional, disordered, nonlinear lattices was performed in \cite{SGF13} for the KG model.
In that study the use of SIs was extended according to the so-called `tangent map method' \cite{SG10_GS11_GES12} to integrate both the orbit itself, as well as a small deviation vector about it, whose evolution is needed
for the computation of a chaos indicator like the maximum Lyapunov exponent (see for example \cite{S10} and references therein). These computations are easier for the KG model than for the DDNLS
system and that is why the former system was chosen in \cite{SGF13}. In that 
paper it was shown that although chaotic dynamics slows down (as is
indicated by the continuous decrease in time of the maximum Lyapunov exponent), it does not cross over into regular dynamics. Nevertheless, performing similar computations also for the DDNLS model is necessary for supporting the possible generality of the results obtained in \cite{SGF13}.

Thus, although our understanding of the dynamical evolution and  the chaotic
behavior of disordered lattices has been improved in recent years, several
important questions remain open:  Will wave packets continue spreading indefinitely, as current numerical simulations indicate \cite{KKFA08,PS08,FKS09,GMS09,SKKF09,SF10,LBKSF10,BLSKF11,SGF13,ABSD14}, or will they eventually exhibit a less chaotic behavior, leading to the halt of spreading, as is conjectured by some researchers \cite{JKA10_A11}?
How does wave packet's chaoticity depend on the initial excitation, and how does it evolve in time?
In order to address these questions we need to perform computationally
expensive numerical simulations and  investigate the asymptotic behavior
of different disordered lattice models. Thus, the
construction of numerical  schemes, which will allow the
accurate and efficient integration of multi-dimensional DDNLS models
is imperative. Several such methods have been proposed and implemented
in recent years, see e.~g.~\cite{SKKF09,BLSKF11,KS01,M13,GESBP13,SGBPE14}.
In this work, we focus our attention to methods
based on symplectic integration techniques and investigate their performance.

The paper is organized as follows. In Sect.~\ref{sec:2part}, after a brief
discussion of the properties of SIs, we describe integration methods that
are based on the splitting of the DDNLS Hamiltonian in two integrable parts.
Section  \ref{sec:3part} is
devoted to symplectic integration techniques  obtained by splitting
the DDNLS Hamiltonian in three
integrable parts. Then,  in Sect.~\ref{sec:num} we compare the performance
of all these numerical schemes for the integration of the DDNLS system, while
in Sect.~\ref{sec:sum} we summarize our results and present our conclusions.

\section{Two part split integration schemes}
\label{sec:2part}

Symplectic integrators are nowadays standard and widely
implemented numerical techniques for Hamiltonian dynamics. Their use
for the integration of the Hamilton equations of motion has two
main advantages: it a) keeps the error of the computed
value of the Hamiltonian function (which is an integral of motion,
usually referred as the system's `energy') bounded, irrespectively of the
total integration time, and b) results in efficient numerical
procedures, as it allows the utilization of relatively large
integration steps $\tau$, which lower the required CPU time. These
characteristics make SIs the ideal tools for the long time
integration of multidimensional DDNLS systems.

Symplectic integrators approximate the real solution of the Hamilton equations of motion
by replacing the dynamics of the Hamiltonian system by appropriately
chosen successive actions of other simpler (and usually integrable)
functions, whose sum is the initial Hamiltonian. Usually,
symplectic splitting methods are implemented by separating
the Hamiltonian in two integrable parts
(for an overview see for example
\cite[Sect.~II.5]{HLW02}\cite{MQ02_MQ06_F06_BCM08} and references therein),
although SIs based on three part splits have also been used
\cite{GESBP13,SGBPE14,K96_C99_GBB08_QPRB10}.

In what follows
we briefly describe the use of SIs for the integration of an autonomous Hamiltonian function which splits in two integrable parts. For this purpose, let us consider a  system of $N$
degrees of freedom ($N$D) described by the Hamiltonian function $H(\vec{x})=h=\mbox{constant}$,
where $\vec{x}= (\vec{q}, \vec{p})$ represents the vector of generalized
coordinates $\vec{q}=(q_1,q_2, \ldots, q_N)$ and momenta $\vec{p}=(p_1,p_2, \ldots, p_N)$. Then, the Hamilton equations of motion are $\frac{d \vec{x}} {dt}= \{\vec{x},H\} = L_H \vec{x}$,
where  $L_H=\{\cdot, H\}$ is a differential operator with $\{\cdot, \cdot \}$
being the Poisson bracket defined by $\{ f,g\}=\sum_{l=1}^{N} \left( \frac{\partial f}{\partial q_l}
\frac{\partial g}{\partial p_l}  - \frac{\partial f}{\partial p_l}
\frac{\partial g}{\partial q_l}\right)$,
for any smooth functions $f(\vec{q},\vec{p})$ and $g(\vec{q},\vec{p})$. The
formal solution of the equations of motion, for initial conditions
$\vec{x}_0=\vec{x}(0)$, is $\vec{x}(\tau)=\sum_{k\geq 0} \frac{\tau^k}{k!} L_H^k \vec{x}_0=e^{\tau L_H}
\vec{x}_0$.

Let us now assume that the Hamiltonian function can be split in two integrable parts as $H(\vec{x})=\mathcal{A}(\vec{x})+\mathcal{B}(\vec{x})$, so
that the action of the operators $e^{\tau L_{\mathcal{A}}}$ and
$e^{\tau L_{\mathcal{B}}}$ is known analytically. Then, a SI of order $n$
approximates the operator $e^{\tau L_H}$ by a product of $p$ operators $e^{a_i \tau L_{\mathcal{A}}}$ and
$e^{b_i \tau L_{\mathcal{B}}}$ (which represent exact integrations over times $a_i \tau$
and $b_i \tau$ of Hamiltonians
$\mathcal{A}(\vec{x})$ and $\mathcal{B}(\vec{x})$  respectively)
according to $e^{\tau L_H}=e^{ \tau \left( L_{\mathcal{A}}+L_{\mathcal{B}} \right)}=\prod_{i=1}^p e^{a_i
  \tau L_{\mathcal{A}}} e^{b_i \tau L_{\mathcal{B}}} + \mathcal{O}(\tau^{n+1})$.
The constants $a_i$ and $b_i$ are chosen specifically to reach the
desired order of the integrator.

The DDNLS system is a one-dimensional lattice model of $N$ coupled
nonlinear oscillators described by the Hamiltonian
\cite{BLSKF11,GESBP13,SGBPE14}
\begin{equation}
H_D=\sum_{i=1}^N \frac{\epsilon_i}{2}\left( q_i^2+p_i^2 \right)+\frac{\beta}{8}
\left( q_i^2+p_i^2 \right)^2- p_{i+1} p_i-q_{i+1}q_i,
\label{eq:HDDNLS}
\end{equation}
where $q_i$ and $p_i$ are, respectively, the generalized position and momentum
of site $i$, the random on-site energy coefficients
$\epsilon_{i}$ are chosen uniformly from the interval
$\left[-W/2, W/2\right]$, with $W$ denoting the
disorder strength, and $\beta \geq 0$ is the nonlinearity's strength,
while fixed boundary conditions ($q_{N+1}=p_{N+1}=0$) are imposed. This
model has two integrals of motion as it conserves the energy $H_D$ (\ref{eq:HDDNLS}) and the
norm
\begin{equation}\label{eq:norm}
    S=\sum_{i=1}^N
\frac{1}{2} \left( q_i^2+p_i^2 \right).
\end{equation}

The Hamiltonian (\ref{eq:HDDNLS}) can be split in two integrable parts
\begin{equation}\label{eq:DDNLS_2split}
    \mathcal{A}= \sum_{i=1}^N \frac{\epsilon_i}{2}\left( q_i^2+p_i^2 \right)+\frac{\beta}{8}
\left( q_i^2+p_i^2 \right)^2 \,\, \mbox{ and } \,\, \mathcal{B}= \sum_{i=1}^N - p_{i+1} p_i-q_{i+1}q_i,
\end{equation}
whose operators $e^{\tau L_{\mathcal{A}}}$,
$e^{\tau L_{\mathcal{B}}}$ can be obtained analytically \cite{com}. In particular, the propagation of  initial conditions
$(q_i, p_i)$ at time $t$, to their final values $(q'_i, p'_i)$ at time
$t+\tau$ is given by the operators
\begin{equation}
e^{\tau L_{\mathcal{A}}}: \left\{ \begin{array}{lll} q'_i & = & q_i \cos(\alpha_i
  \tau)+ p_i \sin(\alpha_i \tau)\\ p'_i & =& p_i \cos(\alpha_i \tau)-
  q_i \sin(\alpha_i \tau) \\
\end{array}\right. , \,\, i=1,2,\ldots, N,
\label{eq:LA}
\end{equation}
\begin{equation}
e^{\tau L_{\mathcal{B}}}: (\vec{q}', \vec{p}')^{\mathrm{T}}= \mathbf{C}(\tau) \cdot (\vec{q}, \vec{p})^{\mathrm{T}},
\label{eq:LB}
\end{equation}
where $\alpha_i=\epsilon_i+\beta(q_i^2+p_i^2)/2$,  $\mathbf{C}(\tau)$
is a matrix
whose expression is obtained in Appendix \ref{sec:ap1}, and ($^{\mathrm{T}}$)
denotes the transpose of a matrix. The operator $e^{\tau L_{\mathcal{A}}}$ has
already been used in the literature \cite{BLSKF11,M13,GESBP13,SGBPE14}, while,
to the best of our knowledge, this is the first time that the expression
(\ref{eq:LB}) of
$e^{\tau L_{\mathcal{B}}}$ with respect to the matrix $\mathbf{C}(\tau)$ is reported. We note that since the integration
step $\tau$ of SIs is typically kept constant, the matrix $\mathbf{C}(\tau)$
appearing in  (\ref{eq:LB}) remains constant for each numerical
implementation of the operator $e^{\tau L_{\mathcal{B}}}$.

Based on this splitting we consider in our study several  SIs of different orders.
An extensively used simple SI of order two is the so-called leap-frog ($LF$) or
Verlet integrator (see for example \cite[Sect.~I.3.1]{HLW02})
    \begin{equation}
    LF(\tau) =
    e^{\frac{\tau}{2} L_{\mathcal{A}}}
    e^{\tau L_{\mathcal{B}}}
    e^{\frac{\tau}{2} L_{\mathcal{A}}}.
    \label{eq:saba1}
    \end{equation}
This is an integrator of
3 steps, i.e.~3 individual applications of operators of integrable
Hamiltonian functions. The $S\mathcal{A}\mathcal{B}\mathcal{A}_2$
integrator \cite{LR01}
    \begin{equation}
    S\mathcal{A}\mathcal{B}\mathcal{A}_2(\tau) =
    e^{c_1 \tau L_{\mathcal{A}}}
    e^{d_1 \tau L_{\mathcal{B}}}
    e^{c_2 \tau L_{\mathcal{A}}}
    e^{d_1 \tau L_{\mathcal{B}}}
    e^{c_1 \tau L_{\mathcal{A}}},
    \label{eq:saba2}
    \end{equation}
with $c_1=\frac{1}{2}\left( 1- \frac{1}{\sqrt{3}} \right)$,
    $c_2=\frac{1}{\sqrt{3}}$, $d_1= \frac{1}{2}$, is another integrator of
    order two having
    5 steps. The order of this integrator can be improved
    when the term
    $\{ \mathcal{B},\{ \mathcal{B},\mathcal{A} \}\}$ leads to an
    integrable system, as in the common
    situation of $\mathcal{A}$ being the usual kinetic energy  and
    $\mathcal{B}$ the potential energy depending only
    on positions. In that case, the addition of two extra corrector steps at
    each end of the integration scheme increases its order
    from two to four. For the DDNLS
    Hamiltonian (\ref{eq:HDDNLS}) we have
    \begin{equation}\label{eq:BA}
    \{ \mathcal{B},\mathcal{A} \} =\sum_{i=1}^N
    \alpha_i \left[q_i (p_{i-1}+p_{i+1}) - p_i(q_{i-1}+q_{i+1}) \right]
    \end{equation}
    and consequently
    \begin{equation}\label{eq:BBA}
    \begin{array}{cl}
      \{ \mathcal{B},\{ \mathcal{B},\mathcal{A} \}\} =
      \displaystyle \sum_{i=1}^N \Bigg\{  &
       \beta \Big[ q_i (p_{i-1}+p_{i+1}) - p_i(q_{i-1}+q_{i+1}) \Big]^2 + \\
       & + \Big( \alpha_i- \alpha_{i-1}\Big)

       \Big[ q_{i-1}^2+  p_{i-1}^2 +q_{i-1}q_{i+1} +p_{i-1}p_{i+1}\Big] +\\
       & + \Big( \alpha_i- \alpha_{i+1}\Big)
       \Big[ q_{i+1}^2+  p_{i+1}^2 +q_{i-1}q_{i+1} +p_{i-1}p_{i+1}\Big]
       \Bigg\},
    \end{array}
    \end{equation}
    which does not seem to be integrable. Thus, we do not combine the
    $S\mathcal{A}\mathcal{B}\mathcal{A}_2$ integrator with a
    corrector term, as was done for example for the KG model \cite{SKKF09}.

We also consider two fourth order integrators: the one found by Forest and Ruth
\cite{FR90} and by Yoshida \cite{Y90}     having 7 steps
    \begin{equation}
    S^4(\tau)=
    e^{c_1 \tau L_{\mathcal{A}}}
    e^{d_1 \tau L_{\mathcal{B}}}
    e^{c_2 \tau L_{\mathcal{A}}}
    e^{d_2 \tau L_{\mathcal{B}}}
    e^{c_2 \tau L_{\mathcal{A}}}
    e^{d_1 \tau L_{\mathcal{B}}}
    e^{c_1 \tau L_{\mathcal{A}}},
    \label{eq:FR}
    \end{equation}
    with $c_1=\frac{1}{2\left(2-2^{1/3}\right)}$,
    $c_2=\frac{1-2^{1/3}}{2\left(2-2^{1/3}\right)}$,
    $d_1= \frac{1}{2-2^{1/3}}$,
    $d_2= -\frac{2^{1/3}}{2-2^{1/3}}$, and  the $\mathcal{A}\mathcal{B}\mathcal{A}864$
      method introduced in
      \cite{BCFLMM13} having 15 steps.  The
      particular values of the coefficients appearing in the
      expression of the $\mathcal{A}\mathcal{B}\mathcal{A}864$ SI are given in Table 3
      of \cite{BCFLMM13}.

    In \cite{Y90} symmetric compositions of  second order
    integrators were used to construct higher order SIs.
    As a sixth order integrator we consider in our study the
    integrator produced by the
    composition method referred as  `solution A' in \cite{Y90}, because
    according to \cite{M95} it shows the best performance
    among the ones presented in \cite{Y90}. According to this technique,
    starting from any second order SI $S^2$, a sixth order
    integrator $S^6$ is constructed as
    \begin{equation}\label{eq:Ycomp6}
    S^6(\tau)=S^2(w_3\tau)S^2(w_2\tau)S^2(w_1\tau)
     S^2(w_0\tau)S^2(w_1\tau)S^2(w_2\tau)S^2(w_3\tau).
    \end{equation}
    The exact values of $w_i$,
    $i=0,1,2,3$ can be found in \cite[Chap.~V, Eq.~(3.11)]{HLW02}
    and \cite{Y90}.
    Using the second order $S\mathcal{A}\mathcal{B}\mathcal{A}_2$ integrator
    (\ref{eq:saba2})
    in (\ref{eq:Ycomp6}) we construct the sixth order
    integrator $S^6_S$ having 29 steps.

\section{Three part split integration schemes}
\label{sec:3part}

In \cite{SGBPE14} several SIs based on a three part split of the
DDNLS Hamiltonian were developed. For these
methods the Hamiltonian (\ref{eq:HDDNLS}) is written as the
sum of the following three integrable parts
\begin{equation}\label{eq:DDNLS_3split}
    A= \sum_{i=1}^N \frac{\epsilon_i}{2}\left( q_i^2+p_i^2 \right)+\frac{\beta}{8}
\left( q_i^2+p_i^2 \right)^2, \,\, B=- \sum_{i=1}^N p_{i+1} p_i\,\,
\mbox{ and } \,\, C= -\sum_{i=1}^N q_{i+1}q_i.
\end{equation}
The operator $e^{\tau L_{A}}$  is exactly the same as operator $e^{\tau L_{\mathcal{A}}}$ (\ref{eq:LA}), while  the operators corresponding to parts $B$ and
$C$ are given by
\begin{equation}
e^{\tau L_B}: \left\{ \begin{array}{lll} p'_i & =& p_i \\ q'_i & = &
q_i-(p_{i-1}+p_{i+1}) \tau \\
\end{array}\right. ,
\label{eq:LBB}
\end{equation}
\begin{equation}
e^{\tau L_C}: \left\{ \begin{array}{lll} q'_i & =& q_i \\ p'_i & = &
p_i+(q_{i-1}+q_{i+1}) \tau \\
\end{array}\right. .
\label{eq:LCC}
\end{equation}
We include in our study the following three part split SIs which showed a particularly good performance in \cite{SGBPE14} for the integration of the DDNLS model: $ABC^4_{[\text{Y}]}$,  $SS^4$, $SS^4_{864}$
 which are of order four, and the sixth order scheme $ABC^6_{[\text{SS}]}$.
We note that schemes $ABC^4_{[\text{Y}]}$ and $ABC^6_{[\text{SS}]}$ are created by
appropriate compositions of the basic three part split integrator of order two
$ABC^2(\tau)=e^{\frac{\tau}{2} L_A} e^{\frac{\tau}{2} L_B}  e^{\tau L_C}
e^{\frac{\tau}{2} L_B} e^{\frac{\tau}{2} L_A}$ \cite{SGBPE14,K96_C99_GBB08_QPRB10},
and have 13 and 45 steps respectively. The other two methods, $SS^4$ and $SS^4_{864}$, are
based on the performance of two successive two part splits of
(\ref{eq:HDDNLS}), i.e.~we implement a two part split SI following  the splitting (\ref{eq:DDNLS_2split})
where the $\mathcal{B}$ part is split again in two parts (the $B$ and $C$
Hamiltonians of (\ref{eq:DDNLS_3split})) and is integrated by the $SABA_2$
integrator.  The specifications of each scheme can
be found in  \cite{SGBPE14}.

\section{Numerical results}
\label{sec:num}

In order to investigate the performance of the various SIs, we consider
as a test case a particular, randomly chosen, disorder realization of (\ref{eq:HDDNLS}) for
$N=1000$ sites, set $W=4$ (i.e.~$\epsilon_i \in [-2,2] $) and $\beta=0.72$, and follow in time the evolution of an initially
homogeneous excitation of the central 21 sites of the lattice. The initial norm
density of the excited sites is set to unity and consequently the wave packet's
norm (\ref{eq:norm}) is $S=S(0)=21$. Initially each excited site gets a
random phase. The particular random configuration we consider in our
simulations results to a total energy $H_D \approx -28.501$. We note that
this configuration corresponds to the so-called `strong chaos' spreading
regime studied in \cite{LBKSF10}.

To evaluate the performance of each tested SI we check whether the obtained
solutions correctly capture the wave packet dynamics by monitoring the
time evolution of the second moment $m_2$ of the wave packet's norm distribution
$z_i=(q_i^2+p_i^2)/(2S)$ \cite{FKS09,SKKF09,LBKSF10,BLSKF11,GESBP13,SGBPE14}.
In addition, we
monitor the preservation of the values of the two integrals $H_D$ (\ref{eq:HDDNLS})
and $S$ (\ref{eq:norm}) by keeping track of the absolute relative errors of the
energy $E_r(t)=|[H_D(t)-H_D(0)]/H_D(0)|$, and the norm
$S_r(t)=|[S(t)-S(0)]/S(0)|$. We also register the
required CPU time $T_c$ needed for performing each simulation.

\begin{figure}[t]
\centering
\includegraphics[scale=1.2, keepaspectratio]{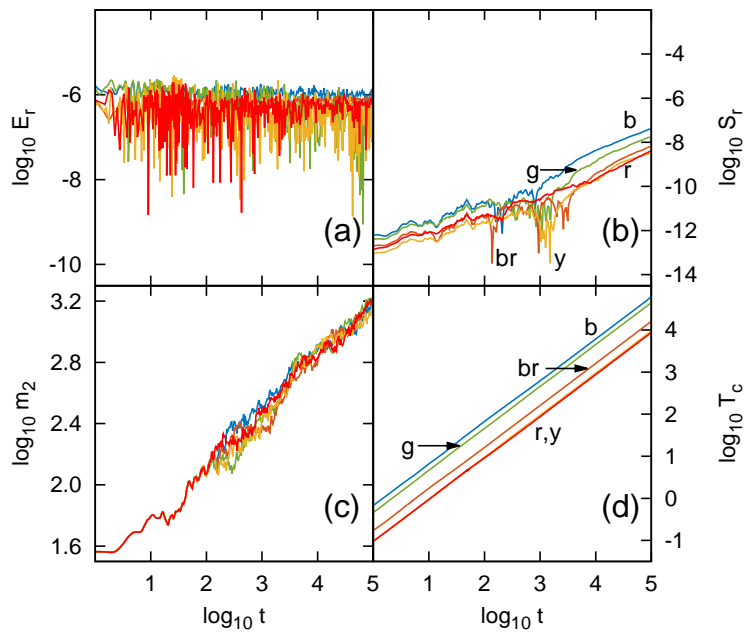}
\caption{(Color online) Results for the integration of
    $H_D$ (\ref{eq:HDDNLS}) by the two part split SIs
    $LF$ for $\tau=0.0025$,
    $S\mathcal{A}\mathcal{B}\mathcal{A}_2$ for $\tau=0.01$,
    $S^4$  for $\tau=0.05$,
    $\mathcal{A}\mathcal{B}\mathcal{A}864$ for $\tau=0.175$ and
    $S^6$ for $\tau=0.25$
    [(b) blue; (g) green; (br) brown; (y) yellow; (r) red]:
    time evolution of the logarithm of (a) the absolute relative energy error $E_r(t)$, (b) the absolute  relative norm error $S_r(t)$ (c)  the second moment $m_2(t)$, and (d)  the required CPU time
    $T_c(t)$ in seconds. Note that in panel (d) the  red and yellow curves practically overlap.}
\label{fig:2}
\end{figure}

In Fig.~\ref{fig:2} we present results obtained by the two part split integrators
discussed in Sect.~\ref{sec:2part}:
$LF$ (\ref{eq:saba1}) [blue curves],
$S\mathcal{A}\mathcal{B}\mathcal{A}_2$ (\ref{eq:saba2}) [green curves],
$S^4$ (\ref{eq:FR}) [brown curves],
$\mathcal{A}\mathcal{B}\mathcal{A}864$ [yellow curves] and
$S^6$ (\ref{eq:Ycomp6}) [red curves].
The integration time step $\tau$ of each method was chosen so that all
schemes keep the relative energy error bounded by practically the same value
$E_r \approx 10^{-6}$  [Fig.~\ref{fig:2}(a)]. Additionally, all these two part split
SIs preserve well the numerical value of the norm $S$ (\ref{eq:norm}), since the corresponding $S_r$ quantities attain relatively small values [Fig.~\ref{fig:2}(b)] (although these values clearly increase in time). This happens because the  norm $S$ (\ref{eq:norm}) is an integral of motion of both the $\mathcal{A}$ and $\mathcal{B}$ Hamiltonians of (\ref{eq:DDNLS_2split}),
as the direct computation of its time derivative in both systems shows ($\frac{dS}{dt}=\{ S,\mathcal{A}\}=0$ and
$\frac{dS}{dt}=\{ S,\mathcal{B}\}=0$). In addition, all integrators
succeed to correctly describe the system's dynamics as they give practically
the same time evolution of $m_2$ [Fig.~\ref{fig:2}(c)]. From the results,
of Fig.~\ref{fig:2}(d) we see that the fourth order
$\mathcal{A}\mathcal{B}\mathcal{A}864$ together with the sixth order scheme $S^6$ (\ref{eq:Ycomp6}) show the best numerical performance as
they both require the least CPU time among all tested methods.

\begin{figure}[t]
\centering
\includegraphics[scale=1.2, keepaspectratio]{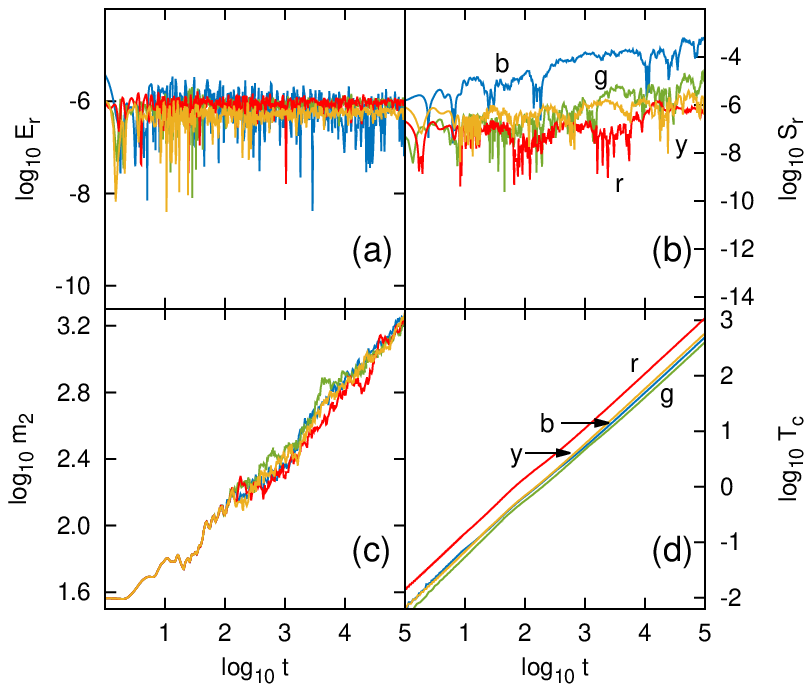}
\caption{(Color online) Results for the integration of
    $H_D$ (\ref{eq:HDDNLS}) by the three part split SIs
    $ABC^4_{[\text{Y}]}$ for $\tau=0.05$,
    $SS^4$ for $\tau=0.05$,
    $SS^4_{864}$  for $\tau=0.125$ and
    $ABC^6_{[\text{SS}]}$ for $\tau=0.225$
    [(b) blue; (r) red; (y) yellow; (g) green].
    The panels are as in Fig.~\ref{fig:2}.
 }
\label{fig:3}
\end{figure}

In Fig.~\ref{fig:3} we see results obtained by the three part
split integrators considered in our study (see Sect.~\ref{sec:3part}):
the fourth order schemes
$ABC^4_{[\text{Y}]}$ [blue curves],
$SS^4$ [red curves],
$SS^4_{864}$ [yellow curves]
and the sixth order integrator $ABC^6_{[\text{SS}]}$  [green curves]. Again,
all integrators keep the relative energy error small ($E_r \lesssim 10^{-6}$)
for the particular choices of the integration steps $\tau$ [Fig.~\ref{fig:3}(a)]
and reproduce correctly the evolution of the wave packet's
second moment  [Fig.~\ref{fig:3}(c)]. A difference with respect to the
two part split SIs of Fig.~\ref{fig:2} is that the values of the relative
norm error $S_r$ [Fig.~\ref{fig:3}(b)] are much larger than the ones reported in Fig.~\ref{fig:2}(b). The reason for this is the fact that the norm $S$ (\ref{eq:norm})
is an integral of motion of $\mathcal{B}$ (\ref{eq:DDNLS_2split})
but not of $B$ and $C$ (\ref{eq:DDNLS_3split}) separately.
Furthermore, we note that also for three part split methods $S_r$ increases in time
although its increase rate is smaller than the one seen in Fig.~\ref{fig:2}(b).
From Fig.~\ref{fig:3}(d) we see that the
sixth order integrator $ABC^6_{[\text{SS}]}$ needs the least CPU time among the tested
SIs. The  $SS^4_{864}$ integrator exhibits the second best behavior, as it requires a slightly larger CPU time than $ABC^6_{[\text{SS}]}$, but  keeps $S_r$ to acceptable levels (smaller than $ABC^4_{[\text{Y}]}$ [Fig.~\ref{fig:3}(b)], with which they require practically the same CPU time [Fig.~\ref{fig:3}(d)]).

It is worth noting that all three part split integrators of
Fig.~\ref{fig:3} require considerably less CPU time than the two part
split schemes of Fig.~\ref{fig:2}, as a direct comparison of
Fig.~\ref{fig:2}(d) and Fig.~\ref{fig:3}(d) reveals, despite the fact
that more operators are involved in the three part split methods.
This happens because the application of
$e^{\tau L_{\mathcal{B}}}$ (\ref{eq:LB}) is computationally very expensive,
especially for
high values of $N$. To illustrate the dependence of the performance
of the integration schemes on the number of the system's degrees of freedom
$N$, we consider the best performing schemes among the two part split methods
(here we used $\mathcal{A}\mathcal{B}\mathcal{A}864$ as preliminary tests showed a slightly better performance than for $S^6$) and the three part split
techniques (i.e.~$ABC^6_{[\text{SS}]}$) and monitor their behavior when
$N$ is varied.

\begin{figure}[t]
\centering
\includegraphics[width=0.48\textwidth, keepaspectratio]{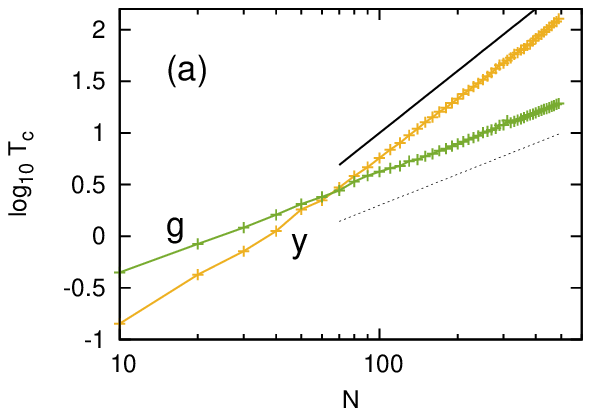}
\includegraphics[width=0.48\textwidth, keepaspectratio]{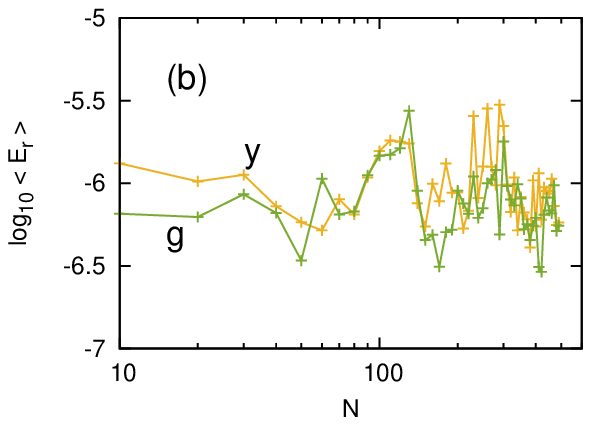}
\caption{(Color online)  The logarithm of (a) the required CPU time $T_c$ and (b)
the
time averaged relative energy error $\langle E_r \rangle$ for the
integration of the DDNLS model (\ref{eq:HDDNLS}) up to the final integration
time $t=10^4$ by the
$\mathcal{A}\mathcal{B}\mathcal{A}864$ [(y) yellow curves] and the
$ABC^6_{[\text{SS}]}$ [(g) green curves] methods for various values of the number of sites $N$.
In (a) the slopes
2 and 1 are denoted by black solid and dotted lines respectively.}
\label{fig:N}
\end{figure}

In Fig.~\ref{fig:N}(a) we report the CPU time $T_c$ needed for each
SI to integrate a lattice of $N$ sites up to $t=10^4$ for various values of $N$ up to $N=500$. For each simulation
the integration time step was kept constant to $\tau=0.175$ for the
$\mathcal{A}\mathcal{B}\mathcal{A}864$ SI and to  $\tau=0.225$ for the $ABC^6_{[\text{SS}]}$ one. The time average (over the whole duration
of each simulation) of the relative energy error $\langle E_r \rangle$
for each method is seen in Fig.~\ref{fig:N}(b). We see that
for the particular choice of integration steps
$\langle E_r \rangle$  does not change significantly as $N$ varies,
remaining close to values around $10^{-6}$ for both methods.

From the results of Fig.~\ref{fig:N}(a) we see that the two part split
SI $\mathcal{A}\mathcal{B}\mathcal{A}864$ requires less CPU time than the
three part split scheme $ABC^6_{[\text{SS}]}$ for values of $N \lesssim 70$.
The difference between the two methods becomes more pronounced as $N$ increases. The
obtained results indicate that $T_c \propto N^2$ for $\mathcal{A}\mathcal{B}\mathcal{A}864$, while   $T_c \propto N$ for $ABC^6_{[\text{SS}]}$, as indicated by the black lines shown in Fig.~\ref{fig:N}(a). This behavior can be easily understood from the different computational complexity of the operators used in these two splitting schemes. While operators $e^{\tau L_B}$(\ref{eq:LBB}) and $e^{\tau L_C}$ (\ref{eq:LCC}) basically require the addition of two vectors, the multiplication of
matrix $\mathbf{C}(\tau)$ with vector $\mathbf x$ in (\ref{eq:LB})  requires $N^2$ operations for each application of $\mathrm e^{\tau L_\mathcal B}$. Therefore, the results of Fig.~\ref{fig:N}(a) clearly show that the three part
split schemes should be used for large lattices, while relatively small lattices ($N$ of the order of a few tens) are better to be integrated by a two part split SI. Let us note, that the behavior shown in Fig.~\ref{fig:N}(a) is independent of the system's spreading regimes as we obtained qualitatively similar results (not
reported here) for initial conditions in the so-called `weak chaos' and
`selftrapping' regimes (see \cite{LBKSF10} for more details on the 
definition of these regimes).

\section{Summary and conclusions}
\label{sec:sum}

The numerical integration of the discrete nonlinear Schr\"odinger equation with disorder is computationally very challenging. In this work we presented and compared various symplectic splitting techniques suitable to perform this task with the required accuracy.

For SIs based on the splitting of the Hamiltonian in two integrable parts, we explicitly presented  (to the best of our knowledge for the first time) an analytical solution for the operator $\mathrm e^{\tau L_\mathcal B}$ (\ref{eq:LB}). Besides being more compact, this form has the advantage that various two part split numerical methods are readily available from the literature. Using several suitable schemes we  integrated the corresponding equations of motion and showed that for this splitting the norm of the wave packet $S$ (\ref{eq:norm}) is preserved naturally to a very high accuracy.

To investigate the importance of these results we also integrated the DDNLS system with several
three part split SIs,
which are already known to be very well suited for this problem
\cite{SGBPE14}. We found  that it is possible to obtain results with the same level of relative energy error much faster with these three part split methods. Furthermore, we showed that (and also explained why) the relative norm error $S_r$ is much larger with respect to the two part schemes.

We also investigated how the performance of these integration schemes depend on the number
of degrees of freedom. For small lattices (with $N \lesssim  70$) the two part split
$\mathcal{A}\mathcal{B}\mathcal{A}864$ scheme proved to be the most efficient one among all tested methods since it required the least CPU time, while at the same time gave very accurate results. Except for very high accuracy needs three part SIs should be used for larger lattices since their computational complexity grows linearly with $N$, and not like $N^2$ as is the case for two part split methods. We note that from the tested three part split integrators the $ABC^6_{[SS]}$ method showed the best performance.

\begin{acknowledgement}
We thank J.~D.~Bodyfelt for sharing his DDNLS computer code at the early stages of this work.
Ch.~S.~would like to thank J.~Laskar for fruitful discussions and for pointing out that the DDNLS model can be split in two integrable parts, as well as the Lohrmann Observatory at the
  Technical University of Dresden for its hospitality during his
visit in 2015, when part of this work was carried
out. Ch.~S.~was supported by the Research Office of
  the University of Cape Town (Research Development Grant, RDG) and by the National Research Foundation of
South Africa (Incentive Funding for Rated Researchers, IFRR).
\end{acknowledgement}

\appendix
\section{Determination of matrix $\mathbf{C}(\tau)$ of equation (\ref{eq:LB})}
\label{sec:ap1}

The equations of motion for the Hamiltonian $\mathcal{B}$ of (\ref{eq:DDNLS_2split}) can be written in the form
\begin{equation}
 \dot{\vec{x}}^{\mathrm{T}} =\mathbf{B}\vec{x}^{\mathrm{T}}= \left(
 \begin{array}{cc}
  \mathbf{0} & \mathbf{A} \\
  \mathbf{-A} & \mathbf{0}\\
 \end{array}
 \right)\vec{x}^{\mathrm{T}},
\label{eq:AB}
\end{equation}
where $\mathbf{A}$, $\mathbf{B}$ are respectively $N\times N$ and  $2N\times 2N$
constant matrices, while $\mathbf{0}$ is the $N\times N$ matrix
having all its elements equal to zero. The $\mathbf{A}$ matrix is a
tridiagonal matrix with all the elements of its main diagonal equal to
zero ($\mathbf{A}_{i,i}=0$, $i=1,2,\ldots,N$), while all the elements of the
first diagonals above and below the main one are equal to $-1$
($\mathbf{A}_{i,i+1}=\mathbf{A}_{i-1,i}=-1$).
The solution of system (\ref{eq:AB}) for a time step $\tau$ is
\begin{equation}
\label{eq:sol_AB}
\vec{x}^{\mathrm{T}}(t+\tau)=e^{\mathbf{B} \tau}\vec{x}^{\mathrm{T}}(t)=
\sum_{k=0}^{\infty} \frac{1}{k!} (\mathbf{B} \tau)^k \vec{x}^{\mathrm{T}}(t)
=\mathbf{C}(\tau)\vec{x}^{\mathrm{T}}(t).
\end{equation}
It can be easily seen by induction  that
\begin{equation}\label{eq:induction}
    \mathbf{B}^{2k}=(-1)^k \left(
 \begin{array}{cc}
  \mathbf{A}^{2k} & \mathbf{0} \\
  \mathbf{0} & \mathbf{A}^{2k}\\
 \end{array}
 \right),\,\,\,\,
\mathbf{B}^{2k+1}=(-1)^k
\left(
 \begin{array}{cc}
  \mathbf{0} & \mathbf{A}^{2k+1} \\
  \mathbf{-A}^{2k+1} & \mathbf{0}\\
 \end{array}
 \right) , \,\,\, k \in \mathbb{N},
\end{equation}
and consequently, matrix $\mathbf{C}(\tau)$ in (\ref{eq:sol_AB}) can be written as
\begin{equation}\label{eq:C_matrix}
    \mathbf{C}(\tau)=e^{\mathbf{B} \tau}=\left(
 \begin{array}{rr}
  \cos (\mathbf{A}\tau) & \sin (\mathbf{A}\tau)  \\
  -\sin (\mathbf{A}\tau) & \cos (\mathbf{A}\tau)\\
 \end{array}
 \right)
\end{equation}
with
\begin{equation}
 \cos(\mathbf{A} \tau) = \sum_{k=0}^\infty \frac{(-1)^k}{(2k)!}\mathbf{A}^{2k}\tau^{2k}
,\,\,\,\,
 \sin(\mathbf{A} \tau) = \sum_{k=0}^\infty \frac{(-1)^k}{(2k+1)!}\mathbf{A}^{2k+1}\tau^{2k+1}.
 \label{eq:cos-sin}
\end{equation}

The evaluation of the elements of matrices $\cos(\mathbf{A} \tau)$
and $\sin(\mathbf{A} \tau)$ can be obtained through the determination
of the eigenvalues and eigenvectors of matrix $\mathbf{A}$ itself (see
for example \cite{S86}). In particular, the eigenvalues $\lambda_k$ of $\mathbf{A}$
are simple and symmetric with respect to zero, and are given by
\begin{equation}
 \lambda_k = -2\cos\left( \frac{k \pi}{N+1} \right), \,\,\,\, k=1,2,\ldots,N,
 \label{eq:eigen}
\end{equation}
while its eigenvectors are orthogonal (since  $\mathbf{A}$ is symmetric) and
have the form
\begin{equation}
	\vec{v}_k = \frac{1}{V_k} \Bigg( \sin \left( \frac{k \pi }{N+1} \right), \sin \left( \frac{2 k \pi }{N+1} \right),  \dots , \sin \left( \frac{ N k \pi }{N+1} \right) \Bigg),
\label{eq:eigenvectors}
\end{equation}
with
\begin{equation}
	V_k^2 = \Vert \vec{v}_k \Vert^2 = \sum_{j=1}^N \sin^2\left( \frac{j k \pi}{N+1} \right)= \frac{2N+1}{4} - \frac{\sin \left( \frac{2N+1}{N+1} k\pi \right)}{4\sin(\frac{k\pi}{N+1})},
\label{eq:Vcoef}
\end{equation}
where $\Vert \cdot \Vert$ denotes the usual Euclidean norm of a vector. The identity
\begin{equation}
	\sum_{j=0}^N \cos \left( j x \right) = \frac{1}{2} \left[ 1 + \frac{\sin \left[ \left( N + \frac{1}{2} \right) x \right]}{\sin \left(\frac{x}{2} \right)} \right]
\label{eq:identity}
\end{equation}	
(see for example \cite[Eq. 1.342-2.]{GR07}) was used to obtain the
last equality of (\ref{eq:Vcoef}).
The matrix  $\mathbf{S}$ having as columns the eigenvectors (\ref{eq:eigenvectors}),
i.e.~$\mathbf{S}=(\vec{v}_1^{\mathrm{T}}, \,\, \vec{v}_2^{\mathrm{T}}, \,\, \ldots, \,\, \vec{v}_N^{\mathrm{T}})$, can be
used to diagonalize  $\mathbf{A}$. Thus, $\mathbf{A}=\mathbf{S}\mathbf{D}\mathbf{S}$ (since $\mathbf{S}^{-1}=\mathbf{S}^{\mathrm{T}}=\mathbf{S}$), with   $\mathbf{D}$
being the diagonal $N\times N$ matrix having as diagonal elements the
eigenvalues  (\ref{eq:eigen}), i.e.~$\mathbf{D}=\mbox{diag}(\lambda_1,
\lambda_2,\ldots,\lambda_N)$. Then, from (\ref{eq:cos-sin}) we see
that matrices $\cos(\mathbf{A} \tau)$
and $\sin(\mathbf{A} \tau)$ defining $\mathbf{C}(\tau)$ in (\ref{eq:C_matrix})
can be written as $\cos(\mathbf{A} \tau)=\mathbf{S}\mathbf{D}_c\mathbf{S}$ and
$\sin(\mathbf{A} \tau)=\mathbf{S}\mathbf{D}_s\mathbf{S}$ with
$\mathbf{D}_c = \mbox{diag}(\cos(\lambda_1 \tau),\cos(\lambda_2 \tau),\ldots,\cos(\lambda_N \tau))$,  $\mathbf{D}_s = \mbox{diag}(\sin(\lambda_1 \tau),\sin(\lambda_2 \tau),\ldots,\sin(\lambda_N \tau))$.
Consequently,  $\mathbf{C}(\tau)$ (\ref{eq:C_matrix}) is a constant
matrix for a fixed integration time step $\tau$.



%
%


\begin{thebibliography}{}
\bibitem{com} We note that in \cite{BLSKF11} it
was incorrectly stated that `\emph{Hamiltonian $\mathcal{B}$  is
not integrable, and thus the operator $e^{\tau L_{\mathcal{B}}}$ cannot be written
explicitly}'.
\end{thebibliography}
\end{document}